\documentclass[runningheads]{llncs}
\usepackage{lipsum}  
\usepackage{multirow}
\usepackage{graphicx}
\usepackage{marvosym}
\usepackage{amsfonts,amssymb}
\usepackage[export]{adjustbox}
\usepackage{pdfpages}

\begin{document}
%
\title{A Conditional Flow Variational Autoencoder for Controllable Synthesis of Virtual Populations of Anatomy}
%
%
\author{Haoran Dou\inst{1} \and
Nishant Ravikumar\inst{1}\thanks{Nishant Ravikumar and Alejandro F. Frangi are joint last authors} \and
Alejandro F. Frangi\inst{1,2,3,4\star}}
\authorrunning{Haoran Dou et al.}
\titlerunning{Flow Autoencoder for Controllable Synthesis of Virtual Population}
%
\institute{Centre for Computational Imaging and Simulation Technologies in Biomedicine (CISTIB), University of Leeds, Leeds, UK \and 
Division of Informatics, Imaging and Data Science, Schools of Computer Science and Health Sciences, University of Manchester, Manchester, UK \and 
Medical Imaging Research Center (MIRC), Electrical Engineering and Cardiovascular Sciences Departments, KU Leuven, Leuven, Belgium \and 
Alan Turing Institute, London, UK
\\
\email{n.ravikumar@leeds.ac.uk, alejandro.frangi@manchester.ac.uk}
}
\maketitle              
\begin{abstract}
The generation of virtual populations (VPs) of anatomy is essential for conducting in silico trials of medical devices. Typically, the generated VP should capture sufficient variability while remaining plausible and should reflect the specific characteristics and demographics of the patients observed in real populations. In several applications, it is desirable to synthesise virtual populations in a \textit{controlled} manner, where relevant covariates are used to conditionally synthesise virtual populations that fit a specific target population/characteristics. 
We propose to equip a conditional variational autoencoder (cVAE) with normalising flows to boost the flexibility and complexity of the approximate posterior learnt, leading to enhanced flexibility for controllable synthesis of VPs of anatomical structures. We demonstrate the performance of our conditional flow VAE using a data set of cardiac left ventricles acquired from 2360 patients, with associated demographic information and clinical measurements (used as covariates/conditional information). 
The results obtained indicate the superiority of the proposed method for conditional synthesis of virtual populations of cardiac left ventricles relative to a cVAE. Conditional synthesis performance was evaluated in terms of generalisation and specificity errors and in terms of the ability to preserve clinically relevant biomarkers in synthesised VPs, that is, the left ventricular blood pool and myocardial volume, relative to the real observed population.
\keywords{Virtual Population  \and Generative Model \and Normalizing Flow.}
\end{abstract}
\section{Introduction}
\textit{In-silico} trials (ISTs) use computational modelling and simulation techniques with virtual twin or patient models of anatomy and physiology to evaluate the safety and efficacy of medical devices virtually~\cite{viceconti2021silico}. Virtual patient populations (VPs), distinct from virtual twin populations, comprise plausible instances of anatomy and physiology that do not represent any specific real patient's data (as in the case of the latter, viz. virtual twins). In other words, VPs comprise synthetic data that help expand/enrich the diversity of anatomical and physiological characteristics that can be investigated within an IST for a given medical device. 
A key aspect of patient recruitment in real clinical trials used to assess device performance and generate regulatory evidence for device approval is the clear definition of inclusion and exclusion criteria for the trial. These criteria define the target patient population considered appropriate/safe to assess the performance of the device of interest. Consequently, it is desirable to enable the \textit{ controlled} synthesis of VPs that may be used for device ISTs, in a manner that emulates the imposition of trial inclusion and exclusion criteria. 

Virtual populations can be considered to be parametric representations of the anatomy sampled from a generative model. Traditional statistical shape models (SSMs), based on methods such as principal component analysis (PCA), have been widely explored in the past decade~\cite{frangi2002automatic,gooya2015bayesian,ravikumar2018group}. Recent studies focus on deep learning-based generative models due to their automatic and powerful hierarchical feature extraction~\cite{bonazzola2021image,dou2022generative}. For instance, Bonazzola~\textit{et al.}~\cite{bonazzola2021image} used a graph convolutional variational auto-encoder (gcVAE) to learn latent representations of 3D left ventricular meshes and used the learnt representations as surrogates for cardiac phenotypes in genome-wide association studies. Dou~\textit{et al.}~\cite{dou2022generative} proposed learning the shape representations of multiple cardiovascular anatomies using gcVAE independently and then assembling them into complete whole-heart anatomies termed virtual heart chimaeras. Other studies have investigated conditional-generative models for synthesis of VPs of anatomies. For example, Beetz~\textit{et al.}~\cite{beetz2022generating} employed a conditional VAE (cVAE), conditioned on gender and cardiac phase, to allow the synthesis of VPs from biventricular anatomies. In subsequent work~\cite{beetz2022multi,li2023deep}, they extended their method to a multidomain VAE to model biventricular anatomies at multiple times (across the cardiac cycle), using patient-specific electrocardiogram (ECG) signals as additional conditioning information (in addition to patient demographic data and standard clinical measurements) to guide the synthesis.  
All aforementioned methods model the latent space in the VAEs/cVAEs as a multivariate Gaussian distribution with a diagonal covariance matrix. This limits the flexibility afforded to the cVAE, as the Gaussian distribution, being unimodal, is a poor approximation to multimodal latent posterior distributions. This in turn limits the overall variability in anatomical shape that can be captured by standard VAEs and cVAEs. 

In this study, we address the limitations of the state-of-the-art conditional generative models used to synthesise VPs of anatomical structures. In particular, we propose a method to relax the constraint on modelling the latent distribution as a unimodal multivariate Gaussian, to boost the flexibility of the generative model, and to enable conditional synthesis of diverse and plausible VPs generation. Recent advances in normalising flows~\cite{rasal2022deep,rezende2015variational,tomczak2016improving} introduce a new solution for this limitation by leveraging a series of invertible parameterized functions to transform the unimodal distribution to a multimodal one. Motivated by this technique, we propose the first conditional flow VAE (parameterised as a graph-convolutional network) for the task of \textit{controllable} synthesis of VPs of anatomy. The contributions are as follows: (i) we introduce normalising flows to learn a multimodal latent posterior distribution by transforming the latent variables from a simple unimodal distribution. This helps the generative model capture greater anatomical variability from the observed real population, leading to the synthesis of more diverse VPs; (ii) we condition the flow-based VAE on patient demographic data and clinical measurements. This enables conditional synthesis of plausible VPs (given relevant covariates/conditioning information as inputs), which reflect the observed correlations between nonimaging patient information and anatomical characteristics in the real population.

\section{Methodology}
In this study, we propose a cVAE model equipped with normalising flows for controllable synthesis of VPs of cardiovascular anatomy. A schematic of the proposed conditional flow VAE network architecture is shown in Figure 1. We employ normalising flows in the latent space of the cVAE to transform the initial Gaussian posterior to a complex multimodal distribution. 

\begin{figure*}[!htbp]
	\centering
	\includegraphics[width=0.9\linewidth]{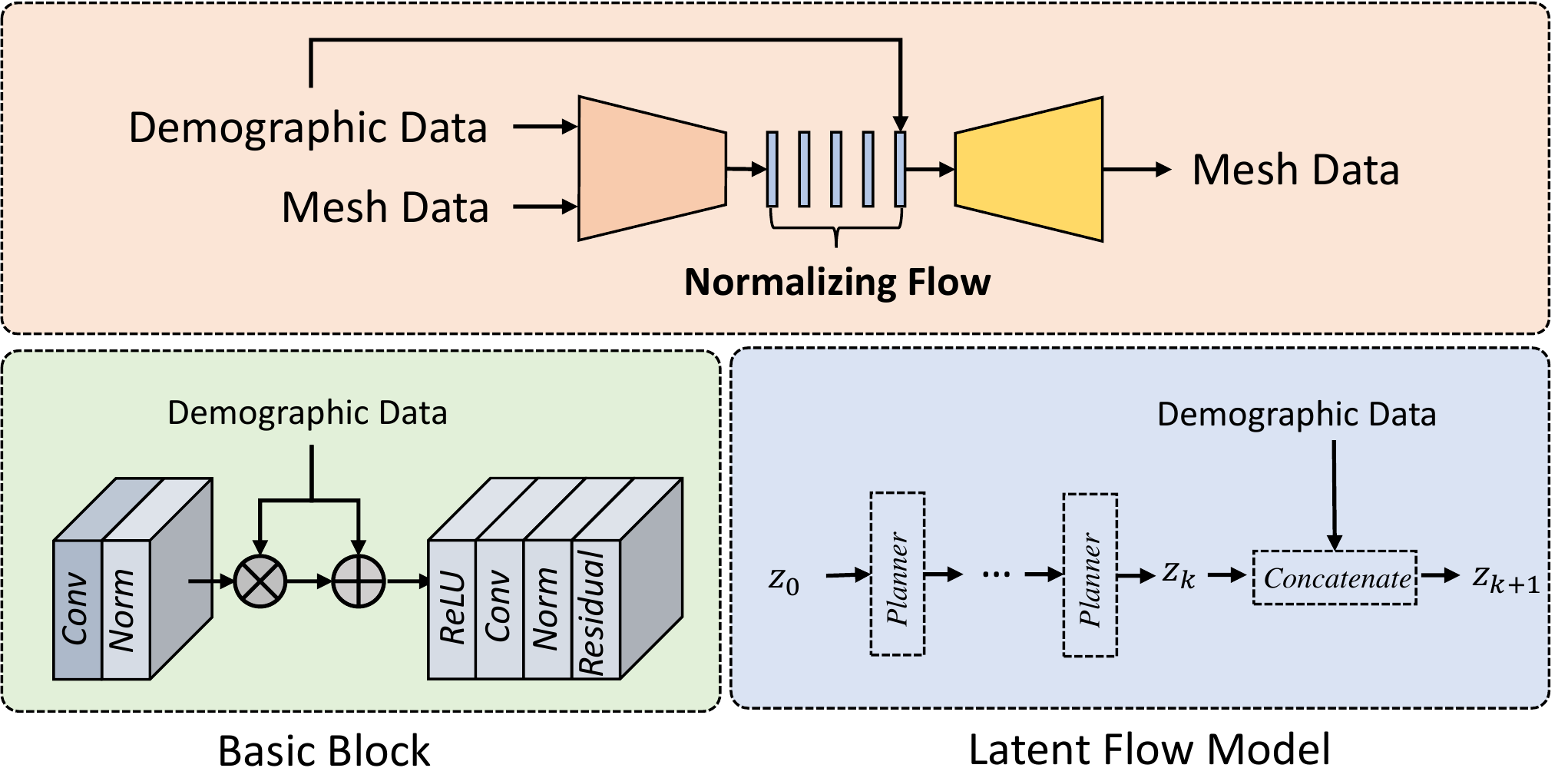}	
	\caption{Schematic illustration of our proposed conditional flow VAE}
	\label{fig:framework}
\end{figure*}

\textbf{Conditional Variational Autoencoder:} A VAE is a probabilistic generative model/network~\cite{kingma2013auto} that comprises an encoder and a decoder network branch. The encoder learns a mapping from the input data to a low-dimensional latent space that abstracts the semantic representations from the observations, and the decoder reconstructs the original data from the low-dimensional latent representation. The latent space from which the observed data is generated is given by approximating the posterior distribution of the latent variables using variational inference. The VAE network is trained by maximising the evidence lower bound (ELBO), which is a summation of the expected log-likelihood of the data and the Kullback-Leibler divergence between the approximate posterior and some assumed prior distribution over the latent variables (typically a multivariate Gaussian distribution). Despite its effectiveness in capturing some of the observed variability in the training population (e.g. of anatomical shapes or images), VAEs do not provide any control over the generation process and hence cannot guarantee that the generated population anatomical shapes are representative of target patient populations with specific inclusion/exclusion criteria. Controllable synthesis of anatomical VPs is essential for constructing meaningful cohorts for use in ISTs. Conditional VAE~\cite{sohn2015learning} is a VAE-variant that uses additional covariates/conditioning information in addition to the input data (e.g. anatomical shapes) to learn a conditional latent posterior distribution (conditioned on the covariates), enabling controllable synthesis of VPs during inference (given relevant covariates/conditioning information as input).

Our conditional flow VAE (cVAE-NF) is a graph-convolutional network which takes as input a triangular surface mesh representation of an anatomical structure of interest, i.e., the Left Ventricle (LV) in this study, and its associated covariates/conditioning variables, i.e., the patient demographic data and clinical measurements, such as gender, age, weight, blood cholesterol, etc., and outputs the reconstructed surface mesh. Each mesh is represented by a list of 3D spatial coordinates of its vertices and an adjacency matrix defining vertex connectivity (i.e. edges of mesh triangles). The encoder and decoder contain five residual graph-convolutional blocks, respectively. Each block comprises two Chebyshev graph convolutions, each of which is followed by batch normalisation and ELU activation. A residual connection is added between the input and the output of each graph-convolutional block. Hierarchical mesh down/up-sampling operations proposed in CoMA~\cite{ranjan2018generating} are adopted after each block to capture the global and local shape context. The VAE model is conditioned on covariates by scaling the hidden representations in the encoder similar to adaptive instance normalization~\cite{huang2017arbitrary} given the covariates as input to generate the scaling factor, and by concatenating the covariates with the latent variables before decoding.

\begin{figure*}[!htbp]
	\centering
	\includegraphics[width=1.0\linewidth]{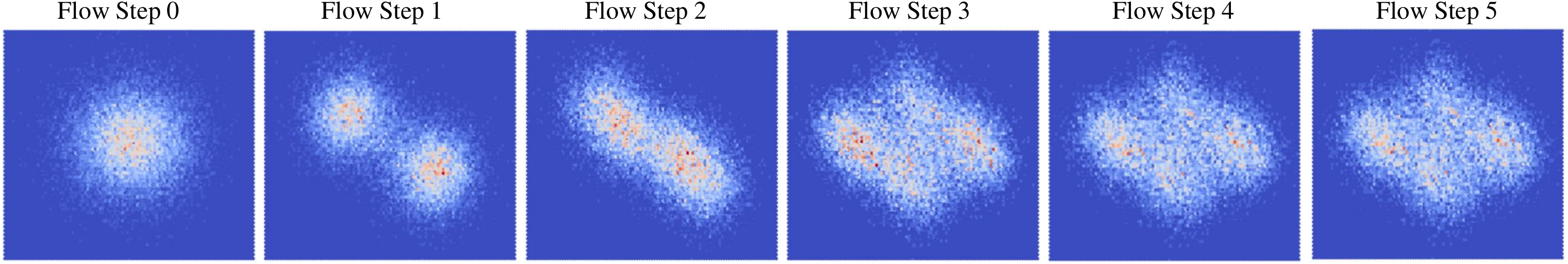}	
	\caption{Effect of normalising flow on Gaussian distribution. Step 0 is the initial two-dimensional Gaussian distribution, and step 1-5 represents the distribution of latent variables transformed by the normalising flow layers (i.e., planar flow).}\label{fig:nf}
\end{figure*}

\textbf{Flexible Posterior using Normalizing Flow:} Vanilla cVAEs model the approximate posterior distribution using Gaussian distributions with a diagonal covariance matrix. However, such a unimodal distribution is a poor approximation of the complex true latent posterior distribution in most real-world applications (e.g. for shapes of the LV observed across a population), limiting the anatomical variability captured by the model. In this study, we introduce normalising flows to construct a flexible multi-modal latent posterior distribution by applying a series of differentiable, invertible/diffeomorphic transformations iteratively to the initial simple unimodal latent distribution. As shown in Fig.~\ref{fig:nf}, a two-dimensional Gaussian distribution can be transformed into a multi-modal distribution by applying several normalising flow steps to the former.

Consider an invertible and smooth mapping function $f: \mathbb{R}^{d} \rightarrow \mathbb{R}^{d}$ with inverse $f^{-1}=g$, and a random variable $\textbf{z}$ with distribution $q(\textbf{z})$. The transformed variable $\textbf{z}^{\prime}=f(\textbf{z})$ follows a distribution given by: 
\begin{equation}
    q( \textbf{z}^{\prime}) = q(\textbf{z}) \left\vert {\rm det} \frac{\partial f}{\partial \textbf{z}}  \right\vert^{-1}
\end{equation}
where the ${\rm det} \frac{\partial f}{\partial \textbf{z}}$ is the Jacobian determinant of $f$. Therefore, we can obtain a complex multi-modal density by composing multiple invertible mappings to transform the initial, simple and tractable density sequentially, as follows,
\begin{equation}
    \textbf{z}_{i} = f_{i}\circ \dots \circ f_{2} \circ f_{1}(\textbf{z}_{0})
\end{equation}
\begin{equation}
\label{equ:lnqz}
    \ln q_{i}(\textbf{z}_{i}) = \ln q_{0}(\textbf{z}_{0}) - \sum^{i} \ln \left\vert {\rm det} \frac{\partial f_{i}}{\partial \textbf{z}_{i-1}} \right\vert
\end{equation}

The specific mathematical formulation of the normalising flow function is important and must be chosen with care to allow for efficient gradient computation during training, scalable inference, and efficiency in computing the determinant of the Jacobian. In this study, we leverage the planar flow in~\cite{rezende2015variational} as a basic unit of our latent normalising flow net. Specifically, each transformation unit is given by,
\begin{equation}
    f(\textbf{z}) = \textbf{z} + \textbf{u} h(w^{\top}\textbf{z} + b)
\end{equation}
where $\textbf{w} \in \mathbb{R}^d$, $\textbf{u} \in \mathbb{R}^d$ and $b\in \mathbb{R}$ are learnable parameters; $h(\cdot)$ is a smooth element-wise non-linear function with derivative $h^{\prime}(\cdot)$ (we use $\tanh$ in our study) and $\mathbf{z}$ denotes the latent variables sampled from the posterior distribution. Therefore, we could compute the log determinant of the Jacobian term in $O(D)$ time as follows:
\begin{equation}
    \phi(\textbf{z}) = h^{\prime}( \textbf{w}^{\top} \textbf{z}+b) \textbf{w}
\end{equation}
\begin{equation}
    \left\vert {\rm det} \frac{\partial f_{i}}{\partial \textbf{z}_{i-1}} \right\vert = \left\vert \rm{det}(\textbf{I} + \textbf{u} \phi(z)^{\top}) \right\vert = \left\vert 1+\textbf{u}^{\top}\phi(\textbf{z}) \right\vert
\end{equation}

Finally, the network is trained by optimizing the modified ELBO based on equation~\ref{equ:lnqz}:
\begin{equation}
    \ln p(\textbf{x} \vert \textbf{c}) \geq \mathbb{E}_{q(\textbf{z}_{0} \vert \textbf{x}, \textbf{c})} \left\lbrack \ln p(\textbf{x} \vert \textbf{z}_{i}, \textbf{c}) + \sum^{i} \ln \left\vert {\rm det} \frac{\partial f_{i}}{\partial \textbf{z}_{i-1}} \right\vert \right\rbrack - {\rm KL}(q(\textbf{z}_{0} \vert \textbf{x}, \textbf{c}) \| p(\textbf{z}_i))
\end{equation}
where, $\ln p(\textbf{x} \vert \textbf{c})$ is the marginal log-likelihood of the observed data $\textbf{x}$ (i.e. here $\mathbf{x}$ represents an LV graph/mesh), conditioned on the covariates of interest (i.e. patient demographics and clinical measurements) $\textbf{c}$; $i$ is the steps of the normalizing flows. $p(\textbf{x}\vert\textbf{z}_{i}, \textbf{c})$ is the likelihood of data parameterised by the decoder network, which reconstructs/predicts $\textbf{x}$ given the latent variables $\textbf{z}_{i}$, transformed by latent (planar) normalising flows, and the conditioning variables $\textbf{c}$; ${\rm KL}(q(\textbf{z}_{0} \vert \textbf{x}) \| p(\textbf{z}_i))$ is the Kullback-Leibler divergence of the approximate posterior initial $q(\textbf{z}_{0} \vert \textbf{x}, \textbf{c})$ from the prior, $p(z)=\mathcal{N}(z \mid 0, I)$.

\section{Experimental setup and  Results}

\textbf{Data:} In this study, we created a cohort of 2360 triangular meshes of the left ventricle (LV) based on a subset of cardiac cine-MR imaging data available from the UK Biobank (UKBB) by registering a cardiac LV atlas mesh~\cite{rodero2021linking} in manual contours (as described in \cite{xia2022automatic}). We randomly split the data set into 422/59/1879 for training, validation, and testing, respectively. All meshes have the same and fixed graph topology, sharing the same edges and faces but differing in the position of vertices; i.e. there is pointwise correspondence across all shapes. We used 14 covariates available for the same subjects in UKBB as conditioning variables for our model, including, gender, age, height, weight, pulse, alcohol drinker status, smoking status, HbA1c, cholesterol, C-reactive protein, glucose, high-density lipoprotein cholesterol (HDL), insulin-like growth factor 1 (IGF-1), and low-density lipoprotein (LDL) cholesterol. These covariates were chosen because they are known cardiovascular risk factors.

\textbf{Implementation Details:} The framework was implemented using PyTorch on a standard PC with a NVIDIA RTX 2080Ti GPU. We trained our model using the AdamW optimizer with an initial learning rate of 1e-3 and batch size of 16 for 1000 epochs. The feature number for each graph convolutional block in the encoder was 16, 32, 32, 64, 64, and in reverse order in the decoder. The latent dimension was set at 16. The down/up-sampling factor was four, and we used a warm-up strategy~\cite{sonderby2016ladder} to the weight of the KL loss to prevent model collapse.

\textbf{Evaluation metrics:} We compared our model (cVAE-NF) with a traditional PCA-based SSM, two generative models without conditioning information including a vanilla VAE and a VAE with normalising flow (VAE-NF) and the vanilla cVAE. Comparison of the vanilla cVAE can also validate the performance of existing approaches~\cite{beetz2022generating,beetz2022multi} because they are built on the cVAE with different covariates and basic units in the network. We evaluated the performance of all methods using three different metrics: 1) the reconstruction error, which evaluates the generalisability of the trained model to reconstruct/represent unseen shapes, using the distance between the reconstructed mesh with the ground truth/original mesh; 2) the specificity error, which measures the anatomical plausibility of the virtual cohorts synthesised, using the distance between the generated meshes and its nearest neighbour in the unseen real population~\cite{davies2009building}; and 3) the variability in the left ventricular volume in the synthesised cohorts, to assess the diversity of the instances generated in terms of a clinically relevant cardiac index. The variability in LV volume was quantified as the standard deviation of the volumes of LV blood pools (BPVols). The Euclidean distance was used to evaluate all three metrics. Additionally, we measured the activity of the latent dimension using the statistic $A = {\rm Cov}_{\textbf{x}}(\mathbb{E}_{\textbf{z}\sim q(\textbf{z} \vert \textbf{x})} [z])$ of the observations $\textbf{x}$~\cite{burda2015importance}. A higher activity score indicates that a given latent dimension can capture greater population-wide shape variability.

\begin{figure*}[h]
	\centering
	\includegraphics[width=1\linewidth]{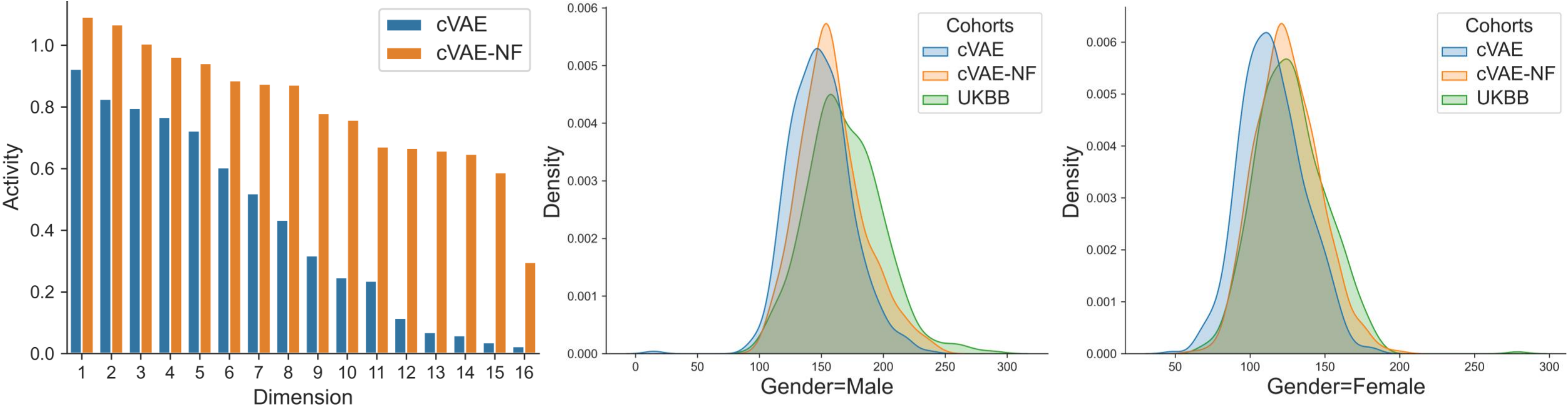}
	\caption{Left: Comparison of the activity scores in different latent dimensions between the cVAE and cVAE-NF; right: Kernel density plots for BPVol from the VPs generated by cVAE and cVAE-NF and the real patient population (UKBB).}\label{fig:activity}
\end{figure*}

\begin{table}[ht]
\centering
\caption{The quantitative results of the investigated methods in a hold-out test dataset. The bold values represent the results are significantly better than those of other methods.}
\label{tab:quantitative}
\begin{tabular}{c|c|c|c}
\hline
\hline
Methods &Reconstruction Error$\downarrow$ &Specificity Error$\downarrow$ &Volume Variability$\uparrow$ \\
\hline
PCA  & \textbf{0.82$\pm$0.16} &1.48$\pm$0.26  &\textbf{32.74}        \\
VAE  & 1.29$\pm$0.21 &\textbf{1.39$\pm$0.98}  &3.00         \\
VAE-NF &0.90$\pm$1.76&1.60$\pm$0.34  &16.03           \\
\hline
cVAE &1.43$\pm$0.26 &\textbf{1.32$\pm$0.21} &28.39      \\
Ours &\textbf{1.23$\pm$0.23} &1.38$\pm$0.20 &\textbf{29.91}      \\
\hline
\hline
\end{tabular}
\end{table}
The results of our method are presented in Table~\ref{tab:quantitative}. Our model outperforms the cVAE in terms of reconstruction error and the amount of volume variability captured in the synthesised VP (the reference volume variability for the real UKBB population was 33.38 $mm^3$). However, the cVAE achieved lower specificity errors than our model. This indicates that our method is better at capturing the population's shape variability, but it also creates some instances that are further away from the real population, resulting in higher specificity errors. We attribute this to the normalising flow's ability to learn a more flexible approximate posterior latent distribution of the observed shapes than the cVAE. This is also seen when comparing the performance of VAE and VAE-NF, where the latter can synthesise significantly more diverse VPs (e.g. it improves the volume variability from 3.00 to 16.03). Figure~\ref{fig:activity} shows the variability captured in each latent dimension. We observe that VAE-NF has higher activity scores in all latent dimensions compared to vanilla cVAE. The normalising flow allows for the approximation of multimodal latent distributions in the generative model, resulting in greater shape variability. Although PCA outperforms our method in terms of generalisation error and volume variability captured, it does not allow for controllable synthesis of VPs based on relevant patient demographic information and clinical measurements, making it less useful for our application of synthesising VPs for use in ISTs.

\begin{figure*}[!tbp]
	\centering
	\includegraphics[width=0.9\linewidth]{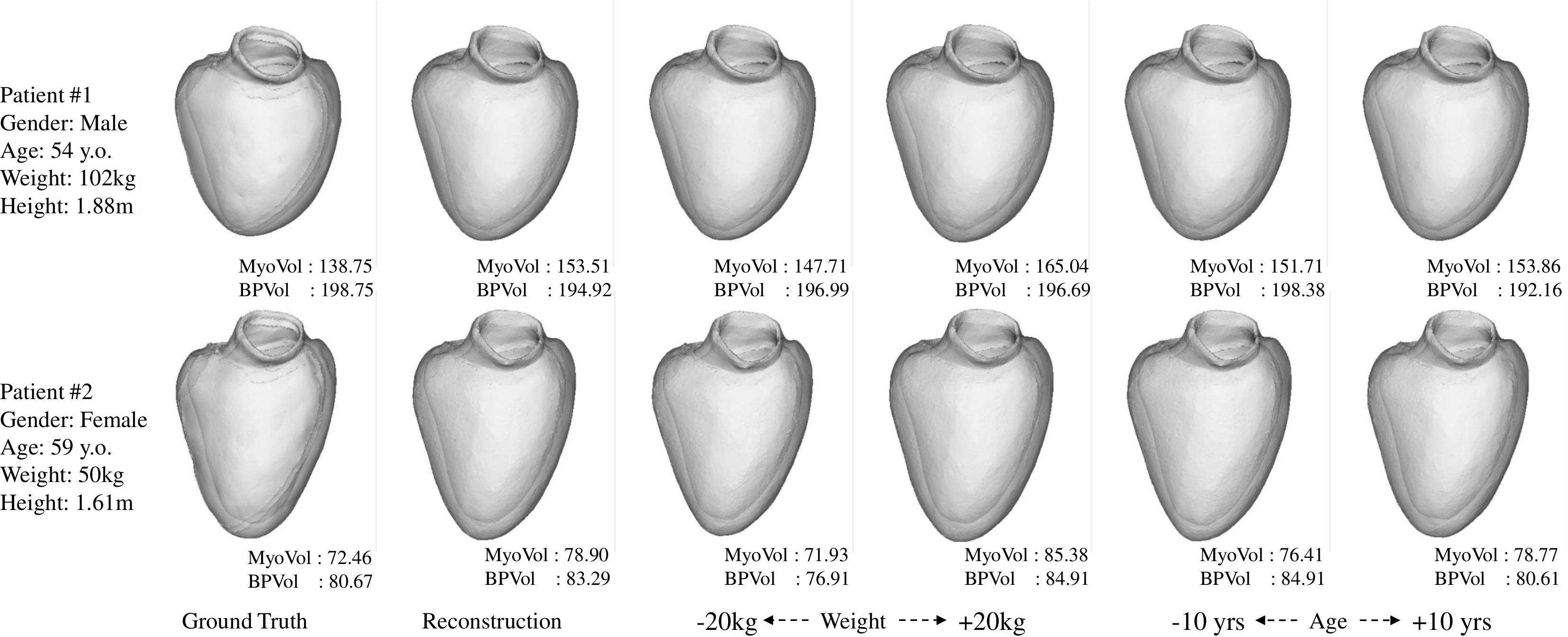}
	\caption{Two representative examples of the reconstructed shapes and their variations through manipulation over two demographic attributes, i.e., weight and Age. MyoVol and BPVol are shown in the bottom right corner.}\label{fig:manipulation}
\end{figure*}

It is essential to capture the distribution of clinically relevant biomarkers (e.g. BPVol) in the synthesised virtual populations (VPs) based on the specified covariates/conditioning information available for real patients, in order to effectively replicate the inclusion/exclusion criteria used during trial design in ISTs. For example, the BPVol of women is known to be lower than that of men~\cite{st2022sex}. To verify this, we generated VPs using cVAE and our method, conditioned on real patient data (covariates) from the UK Biobank. Figure~\ref{fig:activity} summarises the BPVol distribution for both genders in the synthesised VPs and the real UKBB population, and the former accurately reflects the known trend of women having lower BPVol than men. Compared to cVAE, our model generates a VP that more closely matches the distribution of the volume of the LV blood pool observed in the real population. We also visualised the effect of manipulating individual attributes on two real patients in Fig.~\ref{fig:manipulation}. We selected two representative attributes that are significantly associated with BPVol and myocardial volume (MyoVol): weight and age. We observe that BPVol and MyoVol of the LV are positively correlated with the weight of the patients (as expected). On the other hand, increasing the individual's age results in a smaller BPVol, but an increased MyoVol (as visualised in Fig.~\ref{fig:manipulation}), which is known to be due to cardiac hypertrophy caused by aging~\cite{chiao2015aging}.

\section{Conclusion}
We proposed a conditional flow VAE model for the controllable synthesis of VPs of anatomy. Our approach was demonstrated to increase the flexibility of the learnt latent distribution, resulting in VPs that captured greater variability in the LV shape than the vanilla cVAE. Furthermore, our model was able to model the relationship between covariates/conditional variables and the shape of the LV, and synthesise target VPs that fit the desired criteria (in terms of demographics of the patient and clinical measurements) and closely matched the real population in terms of a clinically relevant biomarker (LV BPVol). These results suggest that our approach has potential for the controllable synthesis of diverse, yet plausible, VPs of anatomy. Future work will focus on modelling the whole heart and exploring the impact of individual covariates on VP synthesis in more detail.

\section{Acknowledgement}
This research was carried out using data from the UK Biobank (access application 11350). This work was supported by the Royal Academy of Engineering (INSILEX CiET1819/19), Engineering and Physical Sciences Research Council (EPSRC) UKRI Frontier Research Guarantee Programmes (INSILICO, EP/Y030494/1), and the Royal Society Exchange Programme CROSSLINK  IES$\backslash$NSFC$\backslash$201380.


\bibliographystyle{splncs04}
\bibliography{ref}
%

\includepdfmerge{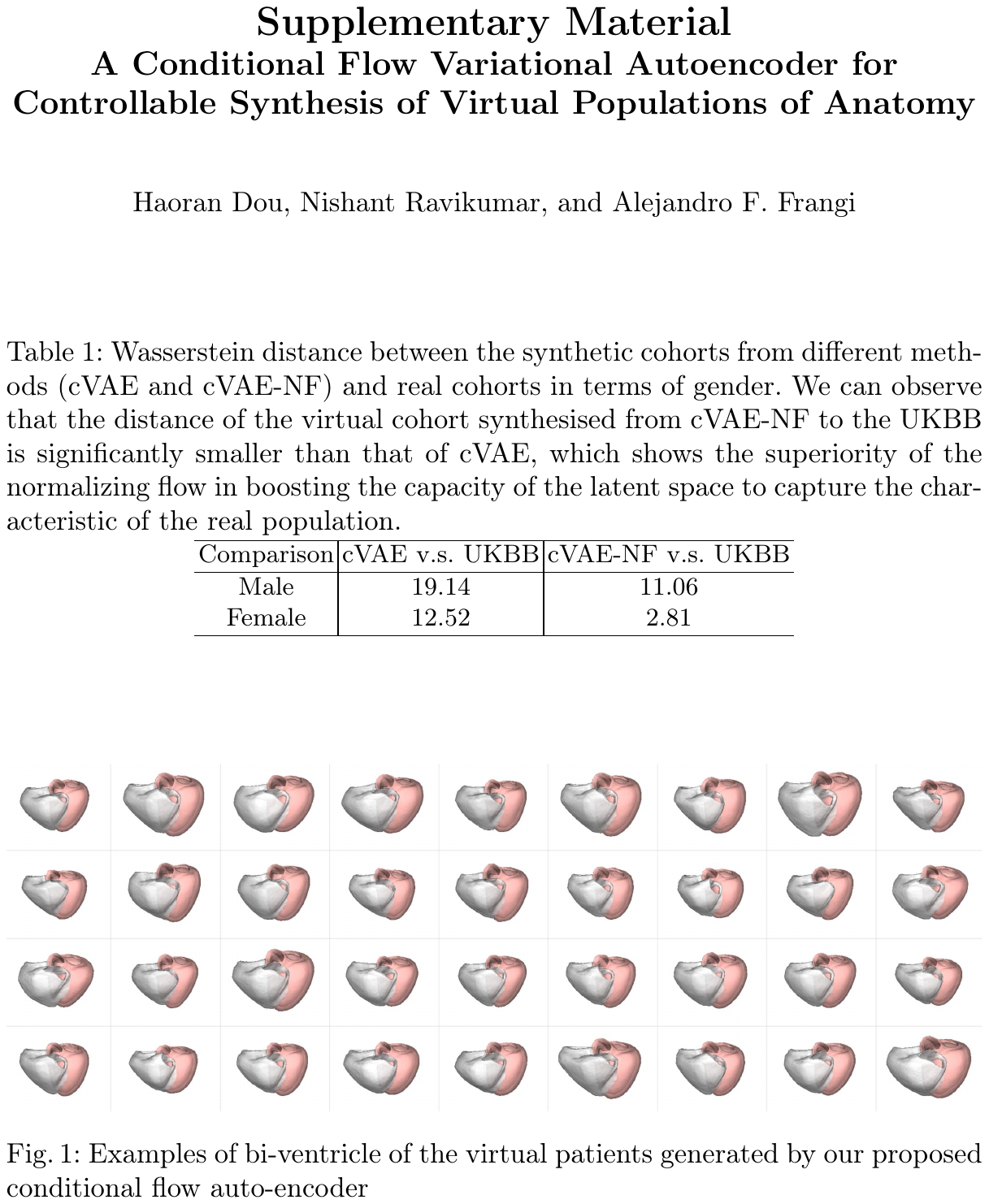, 1}
\includepdfmerge{SupplyMaterial.pdf, 2}
\includepdfmerge{SupplyMaterial.pdf, 3}

\end{document}